\documentclass[aps,prl,twocolumn,superscriptaddress,showpacs]{revtex4-1}
\usepackage{graphicx}
\usepackage{color}
\usepackage{ulem}

\newcommand{\pcsmo}{Pr(Ca$_{0.9}$Sr$_{0.1}$)$_{2}$Mn$_{2}$O$_{7}$}

\begin{document}


\title{Ground State in a Half-Doped Manganite Distinguished by Neutron Spectroscopy}


\author{G. E. Johnstone}

\affiliation{Department of Physics,
        University of Oxford,
        Clarendon Laboratory,
        Oxford, OX1 3PU, United Kingdom}

\author{T. G. Perring}
\email[]{toby.perring@stfc.ac.uk}
\affiliation{ISIS Facility,
        STFC Rutherford Appleton Laboratory,
        Chilton, Didcot OX11 0QX, United Kingdom}
\affiliation{London Centre for Nanotechnology, University College London, London WC1H 0AH, United Kingdom}

\author{O. Sikora}
\affiliation{H. H. Wills Physics Laboratory,
        University of Bristol,
        Bristol BS8 1TL, United Kingdom.}

\author{D. Prabhakaran}
\author{A. T. Boothroyd}
\email[]{a.boothroyd@physics.ox.ac.uk}
\affiliation{Department of Physics,
        University of Oxford,
        Clarendon Laboratory,
        Oxford, OX1 3PU, United Kingdom}


\date{\today}

\begin{abstract}
We have measured the spin wave spectrum of the half-doped bilayer manganite Pr(Ca,Sr)$_{2}$Mn$_{2}$O$_{7}$ in its spin, charge, and orbital ordered phase. The measurements, which extend throughout the Brillouin zone and cover the entire one-magnon spectrum, are compared critically with spin-wave calculations for different models of the electronic ground state. The data are described very well by the Goodenough model, which has weakly interacting ferromagnetic zig-zag chains in the CE-type arrangement. A model that allows ferromagnetic dimers to form within the zigzags is inconsistent with the data. The analysis conclusively rules out the strongly bound dimer (Zener polaron) model.
\end{abstract}



\maketitle




Perovskite manganese oxides and their layered analogues continue to be widely studied not only for the colossal magnetoresistance (CMR) they can show (changes in resistivity of up to several orders of magnitude in magnetic fields of a few Tesla) but also because of the rich physics arising from the coupling of spin, charge, orbital, and lattice degrees of freedom \cite{tokura2006}. The most pronounced CMR phenomena typically arise when the manganite is delicately poised between a ferromagnetic metal and an antiferromagnetic (AFM) charge--orbital (CO) ordered insulator. One of the most ubiquitous AFM CO phases is found in manganites near half doping \cite{wollan1955}, yet despite extensive study motivated by the link with CMR, the electronic state and the mechanisms that stabilize this AFM CO phase remain unclear.

\begin{figure}[h]
\includegraphics[scale=0.5,angle=0,width=0.48\textwidth]{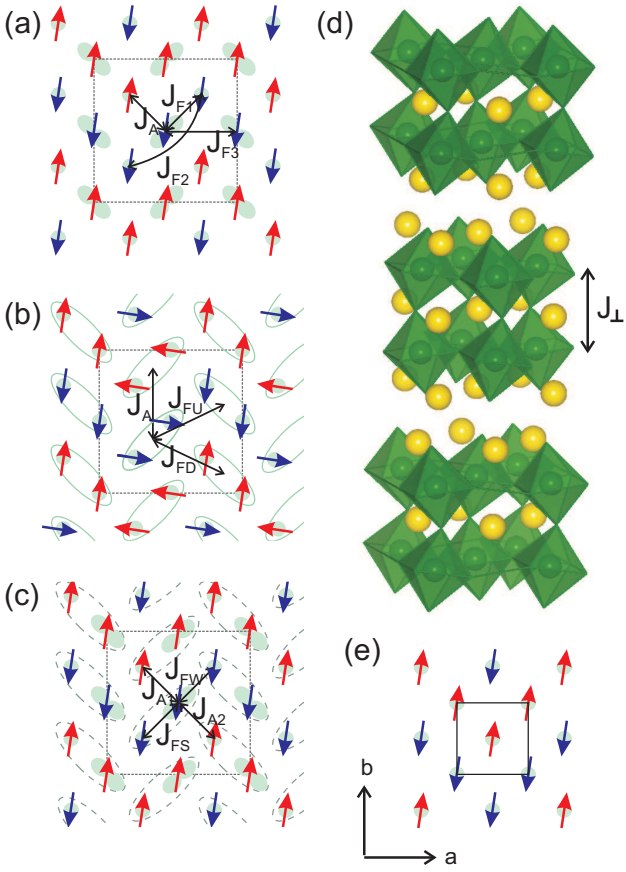}

\centering \caption{\label{fig:struc} (Color online). Magnetic order and exchange interactions within a Mn--O layer for the three models considered here: (a) the Goodenough model with CE-type magnetic order, (b) the Zener polaron model, and (c) the dimer model. The in-plane magnetic unit cells are indicated by the dotted lines. (d) Crystal structure of PCSMO \cite{okuyama2009}, with Mn ions in oxygen octahedra (green) and the Pr, Ca, or Sr ions (yellow) between the Mn--O layers. (e) Projection of the high temperature $Amam$ unit cell onto the Mn--O layer.}
\end{figure}

The first attempt to describe the AFM CO phase was by Goodenough, whose celebrated model is depicted in Fig.~\ref{fig:struc}(a). It is based on the Goodenough-Kanamori-Anderson rules for superexchange \cite{goodenough1955,kanamori1959} and assumes a checkerboard pattern of Mn$^{3+}$ and Mn$^{4+}$ charge order and an associated herringbone pattern of occupied $e_g$ orbitals on Mn$^{3+}$ sites. This results in the CE-type magnetic order, comprising ferromagnetic (FM) zig-zag chains of spins with antiferromagnetic (AFM) alignment to each neighboring chain. Neutron diffraction \cite{radaelli1997,sternlieb1996,jirak2000} and resonant magnetic x-ray scattering \cite{murakami1998,nakamura1999,wilkins2003} studies appear to support this model, but with markedly reduced (to $\lesssim 25$\%) charge disproportionation \cite{radaelli1997,jirak2000,herreromartin2004,goff2004,rodriguez2005,loudon2005}. A distinctly different model, the Zener polaron (ZP) model \cite{daoud-aladine2002,efremov2004}, postulates that pairs of nearest-neighbor Mn ions effectively share an electron through the Zener double-exchange mechanism. These polarons form the bond-centred charge ordering pattern shown in Fig.~\ref{fig:struc}(b), the spins within a polaron being strongly coupled ferromagnetically \cite{zheng2003,ferrari2003,efremov2004,barone2011}. While some structural studies come out strongly against the ZP model \cite{goff2004,grenier2004,rodriguez2005,garcia-fernandez2009}, others support mixed valence or the ZP model \cite{daoud-aladine2002,thomas2004,wu2007}.

The Goodenough and ZP models have the same complex ordering periodicities, and are very difficult to distinguish in diffraction experiments. An alternative approach is to study the spectrum of magnetic excitations, which is determined by the strength of the exchange interactions and will therefore reflect patterns of orbital ordering and dimerisation. In particular, the CE-type and ZP magnetic structures have very different spin-wave spectra \cite{sikora2004}. Previous neutron scattering measurements of the spin-wave dispersion in La$_{0.5}$Sr$_{1.5}$MnO$_4$ are in good agreement with the Goodenough model \cite{senff2006}. However, these measurements neglect the higher energy ($>40$\,meV) part of the spin-wave dispersion, and cannot conclusively rule out the possibility of a Zener polaron state (see below). The same is true of low energy spin-wave data on Nd$_{0.5}$Sr$_{0.5}$MnO$_{3}$ \cite{ulbrich2011}.

Here we report neutron scattering measurements of the complete magnetic spectrum of a half-doped bilayer manganite. We compare the results quantitatively with spin-wave calculations for the Goodenough and ZP models. The results provide a clear preference for the Goodenough model (but with reduced charge disproportionation) over both the ZP model and an alternative model with weaker dimerisation in the FM zigzags.

The AFM CO phase considered here is observed at or near half doping in the Ruddlesden--Popper series \cite{ruddlesden1957} with general formula, $A_{n+1}$Mn$_{n}$O$_{3n+1}$, where $A$ is a rare-earth-metal or alkaline-earth-metal ion. The $n=2$ member of the series, considered here, has a bilayered structure [see Fig.~\ref{fig:struc}(d)]. The magnetic coupling between Mn$_2$O$_4$ bilayers is expected to be $\lesssim 1/100$ of the intra-bilayer interactions \cite{hirota2002}. As well as having a quasi-two-dimensional (quasi-2D) magnetic dispersion, the layered systems have only two crystal twins rather than six as in the three dimensional (3D) manganites ($n=\infty$) \cite{daoud-aladine2002}, which greatly simplifies data analysis.

Our experiment was performed on half-doped \pcsmo\ (PCSMO). PCSMO exhibits charge and orbital order below $T_{\rm{CO}}=370$\,K and antiferromagnetic order below $T_{\rm{N}}=153$\,K  \cite{tokunaga2006,tokunaga2008,Beale2009}. The specific composition of PCSMO used here can be prepared in the form of large crystals suitable for inelastic neutron scattering.

\begin{figure}[t]
\includegraphics[scale=0.5,angle=0,width=0.48\textwidth]{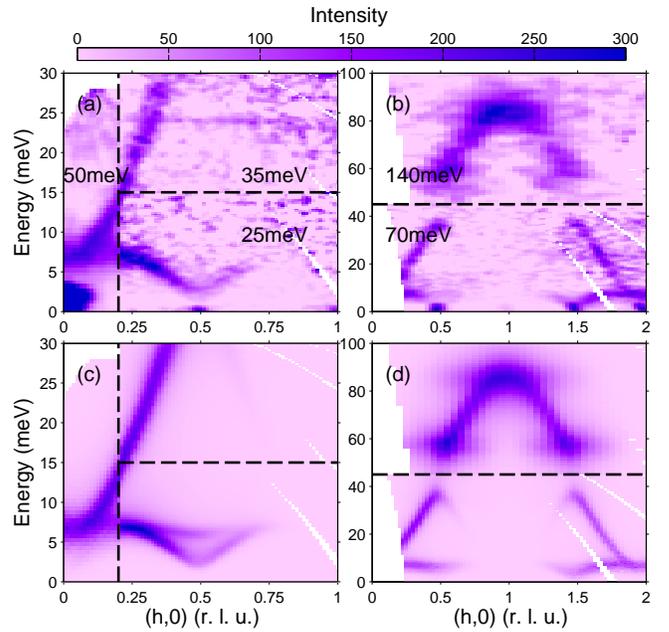}
\centering \caption{\label{fig:cplot} (Color online). Neutron scattering intensity maps of the spin-wave dispersion of PCSMO in the $(h,0)$ direction.  The left-hand plots show energies up to 30\,meV, and the right-hand plots show the whole dispersion. Panels (a) and (b) are experimental data, and panels (c) and (d) are calculations from spin-wave theory for the CE-type order (see text). Regions of the spectrum demarcated by dashed lines were measured with different neutron incident energies and (in some cases) different Brillouin zones. The data have been averaged over a small range of wave vectors in the $(0,k)$ direction. To display the full spectrum with one intensity scale, we have multiplied the data by the monotonic function $f(E)=\frac{E/E_{0}}{1-{\rm e}^{-E/E_0}}$, $E_0=k_{\rm B}\times 10$\,K, which has the effect of suppressing the strong elastic and low energy scattering. Intensity units are mb\,sr$^{-1}$\,meV$^{-1}$ per Mn.}
\end{figure}

Neutron diffraction data confirmed the presence in our sample of two sets of magnetic Bragg peaks below $T_{\rm{N}}$ characteristic of the AFM CO phase magnetic order, one with indices $(0.5, 0, l)$, and the other $(0.5, 0.5, l)$, with $l$ an integer \cite{sternlieb1996}. Throughout this Letter we use the high temperature orthorhombic $Amam$ unit cell \cite{tokunaga2006}, with $a=5.41$\,{\AA}, $b = 5.46$\,{\AA}, and  $c = 19.3$\,{\AA}. In this setting, the $a$ axis is parallel to the CE-type FM zigzags, [see Fig.~\ref{fig:struc}(e)].  Because of twinning, magnetic peaks are also observed at equivalent positions with the $a$ and $b$ axes interchanged.  The diffraction data also showed that the magnetic coupling between layers within a bilayer ($J_{\perp}$) is AFM.

Neutron scattering spectra were recorded at a temperature of 4\,K on a time-of-flight spectrometer in a configuration that made it possible to measure the in-plane magnetic dispersion as a function of neutron energy transfer (see Supplemental Material \cite{Note1}). Figures~\ref{fig:cplot}(a) and 2(b) present intensity maps as a function of energy and wave vector that summarise the important features of the spin waves. Because the dispersion is quasi-2D we give only the in-plane wave vector components. The data have been corrected for an energy-dependent background taken from positions where magnetic scattering is negligible. The magnetic dispersion is plotted in the direction parallel to the CE-type FM zigzags, but also contains features dispersing perpendicular to the zigzags due to twinning.  There is a low energy band emerging from the magnetic Bragg peak at $(0.5,0)$ with a maximum energy of 7\,meV at $(0,0)$, and an upper band which disperses from 7\,meV at $(0,0)$ to a maximum energy of 85\,meV at $(1,0)$. The upper band has an energy gap between 35 and 55\,meV, which extends throughout the Brillouin zone (not all shown). The low energy band is from the direction perpendicular to the FM zigzag, and all of the dispersion over 7\,meV is from the direction parallel the FM zigzag. The periodicity of the upper band, $\Delta h=2$, immediately reveals that the dominant exchange interaction is FM between nearest-neighbor Mn ions projected along the $(h,0)$ direction. The presence of the gap at $h=0.5$ requires a real-space doubling of the period, for example from dimerisation of Mn ions, second-neighbor interactions, or a different moment on alternate Mn ions. A quasi-one-dimensional (quasi-1D) magnetic exchange is implied by the fact that for all energies above $\sim$12 \,meV, the intensity in $(h,k)$ planes at constant energy can be explained by dispersion independent of $k$ once the twinning is accounted for.

\begin{figure}[t]
\includegraphics[scale=0.5,angle=0,width=0.48\textwidth]{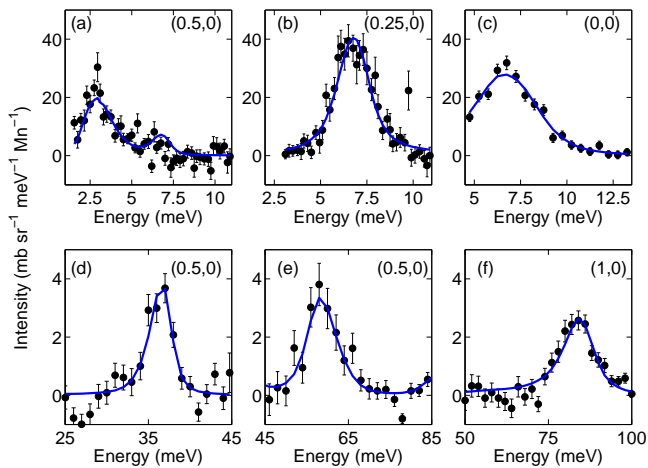}
\centering \caption{\label{fig:cuts} (Color online). Constant wave vector cuts through the experimental data. Fits with resolution corrections are shown by the (blue) lines. A non-magnetic background has been subtracted.}
\end{figure}

Figure~\ref{fig:cuts} shows some examples of energy cuts through the data at different wavevectors. Figures~\ref{fig:cuts}(a)--\ref{fig:cuts}(c) contain data on the lower band, and Figs.~\ref{fig:cuts}(d)--\ref{fig:cuts}(f) on the upper band. The cut in Fig.~\ref{fig:cuts}(a) shows that there is a small energy gap of about 2\,meV at $(0.5,0)$, which is an indication of single-ion anisotropy. Figures \ref{fig:cuts}(d) and \ref{fig:cuts}(e) show the lower and upper limits of the gap in the upper band, and Fig.~\ref{fig:cuts}(f) is a cut through the top of the dispersion at $(1,0)$.

\begin{figure*}[t]
\begin{minipage}{0.4\linewidth}
\centering
\includegraphics[scale=0.5,angle=0,width=0.95\textwidth]{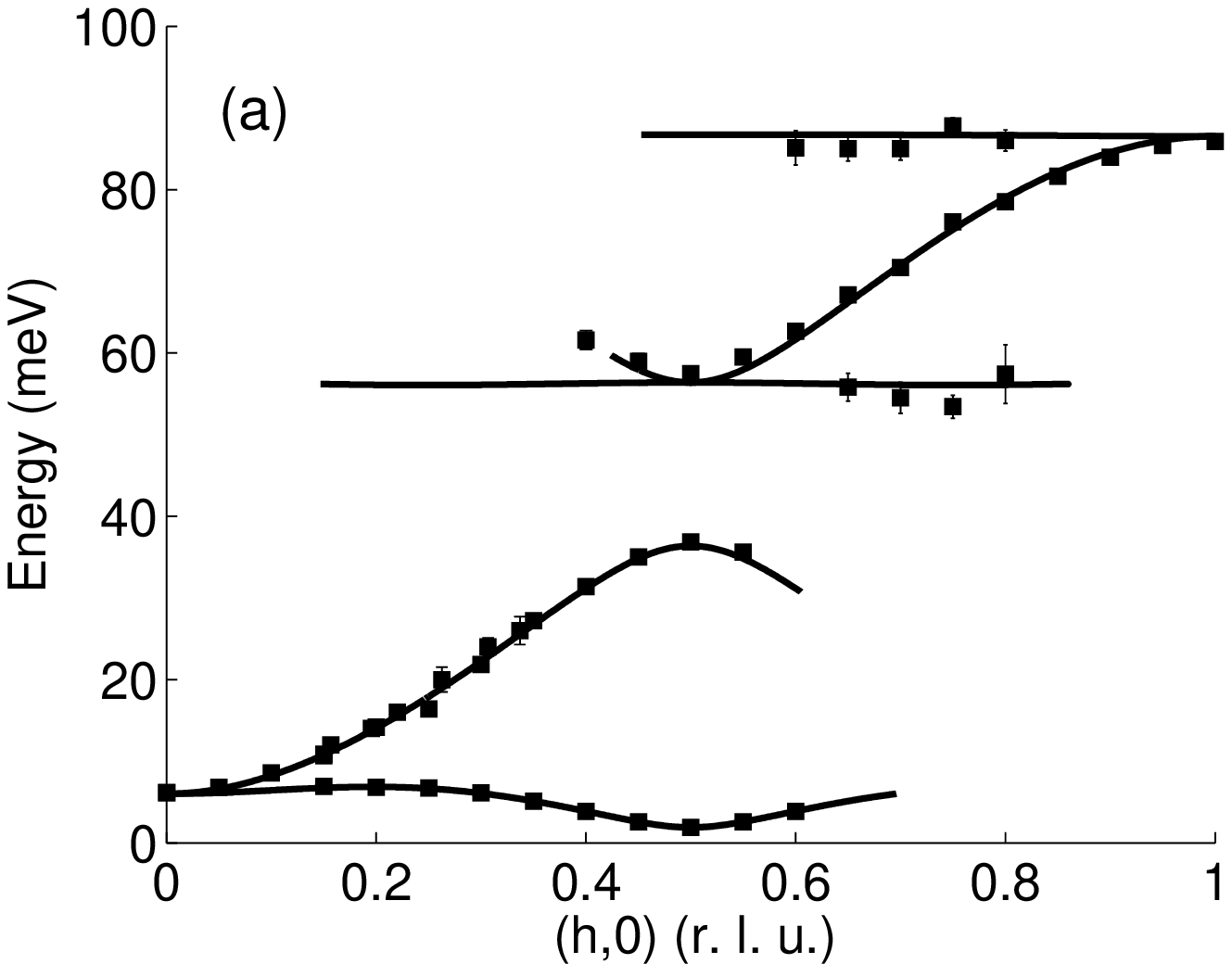}
\end{minipage}%
\begin{minipage}{0.6\linewidth}
\centering
\includegraphics[scale=0.5,angle=0,width=1\textwidth]{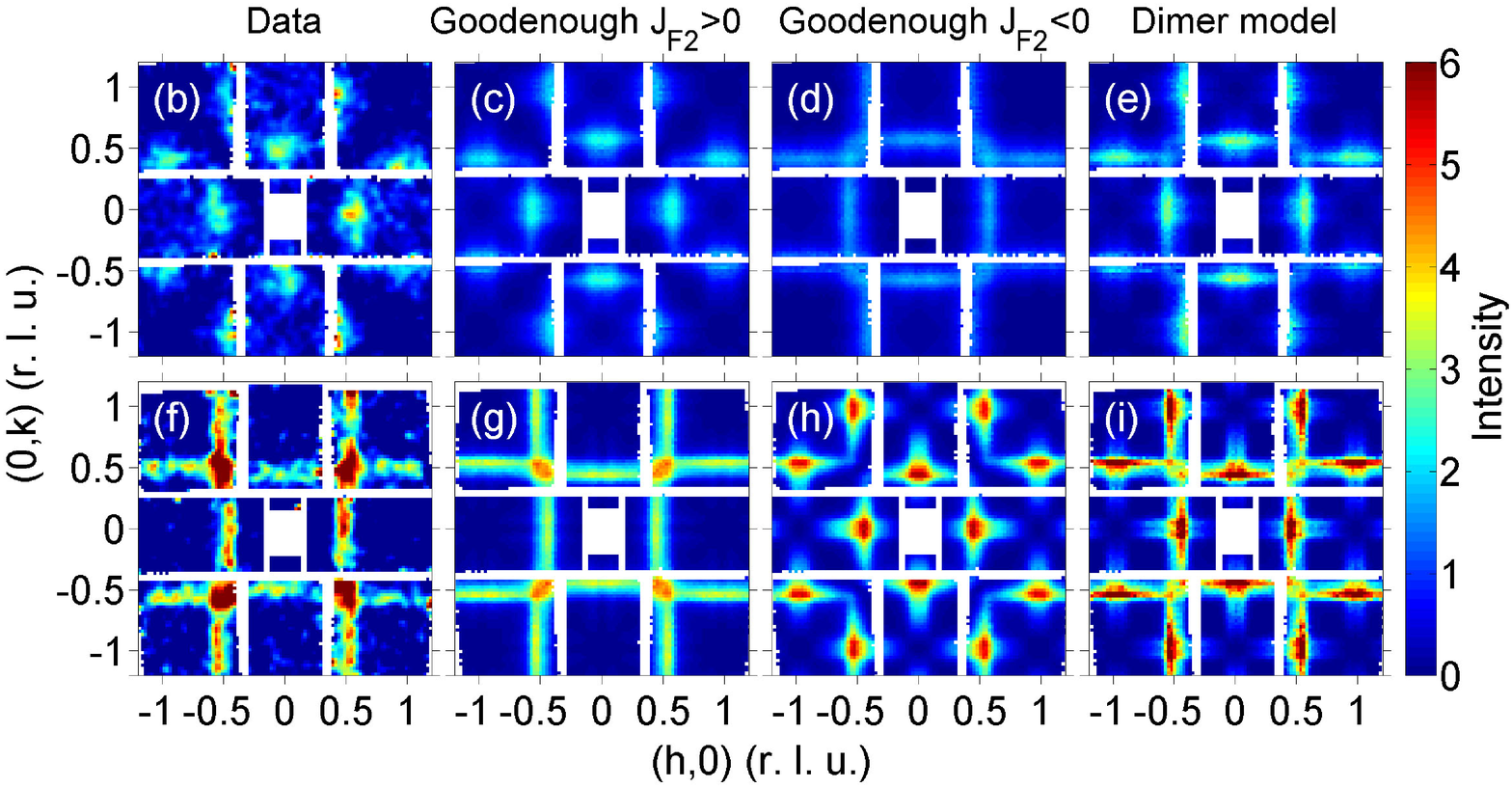}
\end{minipage}
\centering \caption{\label{fig:disp} (Color online). (a) Comparison of the measured magnon dispersion (symbols) parallel to the $(h,0)$ direction with spin-wave calculations for the Goodenough model. (b)--(i) Measured and simulated intensity on $(h,k)$ planes at the upper [(b)--(e)] and lower [(f)--(i)] limits of the gap. The intensity is averaged over 58--62\,meV for the upper limit and 34--36\,meV for the lower limit. Panels (b)--(e) show data, best-fit Goodenough model, best-fit Goodenough model with $J_{\rm F2}$ and $J_{\rm F3}$ exchanged, and best-fit dimer model, respectively. Panels (f)--(i) show the same but for the lower limit. Intensity units are mb\,sr$^{-1}$\,meV$^{-1}$ per Mn.}
\end{figure*}

We evaluated several different models by fitting the calculated magnon dispersion to that determined experimentally, shown in Fig.~\ref{fig:disp}(a) for the direction parallel to $(h,0)$. To obtain the experimental dispersion we analysed the peak positions of approximately 50 cuts like those in Fig.~\ref{fig:cuts}. The spectrometer resolution was included in the analysis (see Supplemental Material \cite{Note1}).

The spin wave models are constructed from the Heisenberg Hamiltonian for the interaction between two spins at sites $i$ and $j$, ${\mathcal H} = -\sum_{\langle i,j\rangle}  J_{ij} {\bf S}_{i} \cdot {\bf S}_{j}$. The spin Hamiltonian for the Goodenough model has five exchange interactions $J_{ij}$, four of which are in the MnO$_2$ plane. These are shown in Fig.~\ref{fig:struc}(a), and are the nearest-neighbor (NN) FM interaction $J_{\rm F1}$, the NN AFM interaction $J_{\rm A}$, and the next-nearest-neighbor (NNN) FM interactions between two Mn$^{4+}$ sites $J_{\rm F2}$ and between two Mn$^{3+}$ sites $J_{\rm F3}$. There is also the NN AFM interaction between layers within a bilayer $J_{\perp}$. Single-ion anisotropy is included by adding $-D \sum_{i} (S_{i}^{z})^{2}$ to ${\mathcal H}$. Initially we will not distinguish between Mn$^{3+}$ and Mn$^{4+}$ and assume the spins on all Mn sites have the same value. The neutron scattering intensity was calculated by linear spin-wave theory.

Spin wave dispersion curves computed from the best-fit Goodenough model are drawn in Fig.~\ref{fig:disp}(a), and Figs.~\ref{fig:cplot}(c) and \ref{fig:cplot}(d) show simulated intensity maps. Taking an average Mn$^{3+}$ and Mn$^{4+}$ spin of $S=7/4$, we obtain fitted parameters $J_{\rm F1}=11.39 \pm 0.05$\,meV, $J_{\rm A}= -1.50 \pm 0.02$\,meV, $J_{\rm F2}= 1.35 \pm 0.07$\,meV, $J_{\rm F3}= -1.50 \pm 0.05$\,meV, $J_{\perp}= -0.88 \pm 0.03$\,meV and $D =0.074 \pm 0.001$\,meV. As there are many spins in the magnetic unit cell, there are many branches in the dispersion diagram. To aid the interpretation of Fig.~\ref{fig:disp}(a), we only plot modes whose calculated intensities are $\geq$10\% of the maximum intensity.

The striking feature of this fit is the dominance of the NN FM interaction $J_{\rm F1}$, almost an order of magnitude larger than any other exchange. It is stronger than the FM exchange in metallic FM manganites, for which $J_{\rm F1} \approx 3.5$--$5.5$\,meV \cite{zhang2007,hirota2002}, and similar to $J_{\rm F1}=9.98$\,meV reported for the AFM CO phase of La$_{0.5}$Sr$_{1.5}$MnO$_4$ \cite{senff2006}. Another notable result is that the NNN interactions between FM aligned spins ($J_{\rm F2}$ and $J_{\rm F3}$) are almost identical in magnitude but opposite in sign, a feature imposed by the energy of the center of the gap and its size. This suggests some frustration in the ground state. The dispersion is unaltered if the values of $J_{\rm F2}$ and $J_{\rm F3}$ are exchanged. However, a comparison of the measured intensity with simulations in $(h,k)$ planes at 60 and 35\,meV [see Figs.~\ref{fig:disp}(b)--\ref{fig:disp}(d) and Figs.~\ref{fig:disp}(f)--\ref{fig:disp}(h), respectively] unambiguously determines that $J_{\rm F2}$ is FM and $J_{\rm F3}$ is AFM.

In the model just described, the gap in the upper band is caused by NNN interactions. Another possibility is that it could arise from alternating spin values due to charge disproportionation. Assuming that the number of $d$ electrons is proportional to the total spin, with $S=2$ associated with Mn$^{3+}$ and $S=3/2$ with Mn$^{4+}$, we have found that charge states of Mn$^{2.68(2)+}$ and Mn$^{4.32(2)+}$ are required to reproduce the observed gap. In contrast, bond valence sums yield Mn$^{3.53(2)+}$ and Mn$^{3.93(2)+}$ \cite{okuyama2009}, a disproportionation only 25\% of that required. This possibility, therefore, can be dismissed.

\begin{figure}[t]
\includegraphics[scale=0.5,angle=0,width=0.48\textwidth]{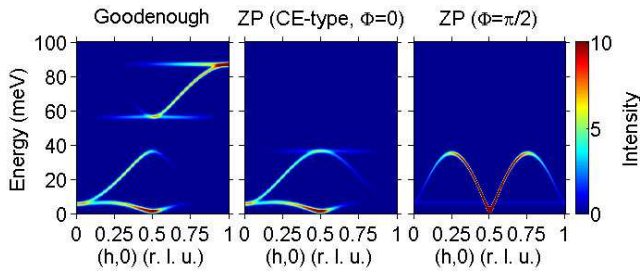}
\centering \caption{\label{fig:lowerband} (Color online). Simulated spectra for the best-fit Goodenough model, the CE-type ZP model (which is equivalent to the strongly-bound dimer model), and the $90 ^\circ$ ZP model.}
\end{figure}

A similar analysis was performed for the ZP model, except the spins in the Hamiltonian refer to the Zener polaron, rather than an individual Mn ion. We used an effective Heisenberg Hamiltonian to capture the essential features of the ZP model \cite{sikora2004}. It has only three in-plane magnetic interactions [see Fig.~\ref{fig:struc}(b)]: the NN AFM interaction between two polarons ($J_{\rm A}$), and two NN interactions reflecting the zig-zag structure ($J_{\rm FU}$ and $J_{\rm FD}$). The out-of-plane exchange $J_{\perp}$ is also included. The magnetic ground state is CE-type (i.e.~the ZP moments are parallel along a zigzag) if $J_{\rm FU}>J_{\rm FD}$ and $|J_{\rm A}|/J_{\rm FU} > J_{\rm FD}/(J_{\rm FU}-J_{\rm FD})$, and the ground state has the $90 ^\circ$ structure shown in Fig.~\ref{fig:struc}(b) when $J_{\rm FU}=J_{\rm FD} \ll |J_{\rm A}|$. These correspond to the $\Phi=0$ and $\Phi=\pi/2$ ground states of Ref.~\cite{efremov2004}, respectively. In neither case can values of the exchange parameters be chosen to to reproduce the observed 35--55\,meV gap or dispersion maximum at $h=1$. To illustrate this, Fig.~\ref{fig:lowerband} shows the spectrum of the best-fit Goodenough model alongside representative simulations for the two ZP models. In the CE-type ZP order, the dispersion parallel to the zigzags has maxima at $h=1/2, 3/2, ...,$ i.e., inconsistent with the data above the gap. In the $90 ^\circ$ ZP phase, the quasi-1D dispersion is perpendicular to the FM zigzags with maxima at $k=1/4, 3/4, ...,$ which because of twinning would also occur at these values of $h$, also inconsistent with the data. Furthermore, the weaker dispersion measured perpendicular to the zigzags confirms that $J_{\rm A}$ is not the dominant exchange parameter.


A third model was tested, which is like the Goodenough model but which allows for dimerization within the zig-zag FM chains. Mn ions are paired similar to the ZP model, but the model does not enforce the rigid coupling between the spin pairs. This dimer model is shown in Fig.~\ref{fig:struc}(c). The important interactions are the two NN FM interactions $J_{\rm FS}$ and $J_{\rm FW}$, the NN AFM interactions $J_{\rm A1}$ and $J_{\rm A2}$, and the intrabilayer interaction $J_{\perp}$. As before, we assume all Mn spins are $S=7/4$, and we set $J_{\rm A1}=J_{\rm A2} \equiv J_{\rm A}$. The fit of this model to the measured dispersion differs from the Goodenough model fit by $\lesssim 2$\% and therefore is practically indistinguishable. The fitted parameters are $J_{\rm FS}= 14.2 \pm 0.08$\,meV, $J_{\rm FW}= 8.43 \pm 0.06$\,meV, $J_{\rm A}= -1.52 \pm 0.01$\,meV, $J_{\perp}= -0.92 \pm 0.03$\,meV, and $D=0.073 \pm 0.001$\,meV. In the dimer model the gap arises from the alternating NN FM exchange without the need for NNN interactions. The dimer model is superficially appealing because of this simplicity and because the value of the intra-dimer exchange $J_{\rm FS}$ appears not unrealistic for double exchange (see above). In fact, the CE-type ZP dispersion corresponds to the limiting case of the dimer model as $J_{\rm FS} \rightarrow \infty$. However, a comparison of the measured intensity with simulations in $(h,k)$ planes at the upper and lower limits of the gap [see Figs.~\ref{fig:disp}(b) and \ref{fig:disp}(e), and Figs.~\ref{fig:disp}(f) and \ref{fig:disp}(i)] rules out the dimer model.

We emphasize that our study can conclusively distinguish between different models only because it includes data above the 35--55\,meV gap. This is demonstrated in Fig.~\ref{fig:lowerband}, in which the spectrum for the dimer model in the strongly bound dimer limit $J_{\rm FS} \gg J_{\rm FW}$ (equivalent to the CE-type ZP model) has been calculated with parameters adjusted to match it to the lower spin-wave branch of the Goodenough model. Either model would be acceptable if the data were restricted to energies below the gap, but only the Goodenough model is consistent with the branches observed above 40\,meV.


This study of the spin wave spectrum of PCSMO clearly demonstrates that the Zener polaron model is not a suitable description of the ground state in the AFM CO phase. The dimer model, which does not impose a rigid coupling between the two Mn ions in a Zener polaron, similarly fails to describe the data.  What distinguishes our data from previously published data on AFM CO phase manganites \cite{senff2006,ulbrich2011} is that they cover the entire one-magnon spectrum, which was essential to rule out the ZP and dimer models conclusively. In contrast, the results are very well described with the original Goodenough model, even without any significant charge ordering. The entire magnetic spectrum can be explained by a single value of the NN FM interactions between Mn sites along the FM zigzags which is larger than that typically found in metallic manganites, strongly suggesting its origin in double exchange. Significant NNN interactions along the zigzags further suggest the $e_g$ electrons are not fully localised.

\begin{acknowledgments}
G.E.J. is grateful to the EPSRC and STFC for the provision of a studentship. O.S. acknowledges financial support from the EPSRC and from the Okinawa Institute of Science and Technology.
\end{acknowledgments}


\end{document}